# Do Users Focus on the Correct Cues to Differentiate Between Phishing and Genuine Emails?

**Kathryn Parsons**
Defence Science & Technology Group
Edinburgh, SA, Australia
Email: Kathryn.Parsons@dsto.defence.gov.au

**Marcus Butavicius**
Defence Science & Technology Group
Edinburgh, SA, Australia
Email: Marcus.Butavicius@dsto.defence.gov.au

**Malcolm Pattinson**
Business School
The University of Adelaide
Adelaide, Australia
Email: Malcolm.Pattinson@adelaide.edu.au

**Agata McCormac**
Defence Science & Technology Group
Edinburgh, SA, Australia
Email: Agata.McCormac@dsto.defence.gov.au

**Dragana Calic**
Defence Science & Technology Group
Edinburgh, SA, Australia
Email: Dragana.Calic@dsto.defence.gov.au

**Cate Jerram**
Business School
The University of Adelaide
Adelaide, Australia
Email: Cate.Jerram@adelaide.edu.au

## Abstract

This paper examines the cues that typically differentiate phishing emails from genuine emails. The research is conducted in two stages. In the first stage, we identify the cues that *actually* differentiate between phishing and genuine emails. These are the consistency and personalisation of the message, the perceived legitimacy of links and sender, and the presence of spelling or grammatical irregularities. In the second stage, we identify the cues that participants use to differentiate between phishing and genuine emails. This revealed that participants often use cues that are not good indicators of whether an email is phishing or genuine. This includes the presence of legal disclaimers, the quality of visual presentation, and the positive consequences emphasised in the email. This study has implications for education and training and provides a basis for the design and development of targeted and more relevant training and risk communication strategies.

**Keywords**

phishing, cyber security, human factors, decision making, cyber threat

## 1 Introduction

Phishing is a term used to describe the attempt to acquire sensitive or personal information, typically by sending an email in which the sender acts as a trustworthy entity. Despite the fact that security researchers and practitioners have been aware of the dangers associated with phishing for almost a decade, it still poses a significant problem today (Furnell 2013), and current technical phishing countermeasures are unable to prevent vulnerable users from being deceived (Purkait 2012).

Previous phishing studies have taken various forms and have measured participants' responses to phishing emails in different ways. Some phishing studies have been conducted 'in-the-wild', where uninformed students received emails that claimed to be from their university. These studies have shown that susceptibility is high, particularly when students were influenced by authority or social context (Ferguson 2005; Jagatic et al. 2007). This approach has also been used within Small and Medium Enterprises and Multi-national Corporations to test their employees (Osterman Research Inc. 2015). Other studies have required participants to make judgements concerning the authenticity of emails or websites, and have shown that many participants have trouble identifying phishing (Dhamija et al. 2006; Furnell 2007; Pattinson et al. 2012). Some of these studies have included qualitative data on the cues of an email that influenced participants; generally speaking, people are more likely to trust emails that are visually appealing, without language errors, and without excessive urgency (Furnell 2007; Jakobsson et al. 2007; Parsons et al. 2013). Users often ignore standard security indicators (Dhamija et al. 2006).

Although the results from previous research suggest that participants often make legitimacy decisions based on the presence of a variety of cues within emails, there has been no rigorous evaluation of which cues are more common in phishing emails and which cues can differentiate between phishing and genuine emails. Some studies (e.g., Olivo et al. 2013) have described the technical features of phishing emails, which usually involve aspects of the email header and URL, but these studies have not examined which email features are clues to identifying a phishing email. Research of this nature could identify which cues are good but poorly understood predictors of phishing. As a consequence, this could then form the basis of phishing education and training. Phishing education has received relatively little research attention (Arachchilage et al. 2013; Arachchilage et al. 2014; Purkait 2012).

## 1.1 Aim of the research

The current study seeks to determine if the cues that participants use to decide on the legitimacy of an email are good indicators of whether an email is phishing or genuine. The structure of this paper is as follows. The next section outlines previous research, including studies involving the categorisation of email cues in Section 2.1, and studies involving the cues identified by participants in previous studies in Section 2.2. This is followed by an explanation of Stage One of our research, in which we determine the cues that can *actually* differentiate between phishing and genuine emails, then Stage Two of our research, in which we determine the cues that participants used to differentiate between phishing and genuine emails. Finally, the significance and real-world implications of these findings are discussed, and conclusions presented.

# 2 Previous Research

## 2.1 Categorisation of email cues

Our review of previous research identified very few studies that attempted to categorise the different cues of emails. Kim et al. (2013) attempted to understand the persuasive cues in phishing emails by categorising 285 phishing emails based on message presentation and content. This included aspects of source credibility, such as the sender's email address, contact methods and company logo, and aspects of argument quality, such as the inclusion of rational, emotional and motivational appeals, as well as time pressure. They concluded that most phishing emails include rational, emotional or motivational appeals, but many do not include contact methods for the sender (Kim et al. 2013).

Similarly, Blythe et al. (2011) collected 100 emails from a phishing archive and conducted a content analysis on the purported sender and premise of the email, the presence of logos, and the number of spelling and grammatical mistakes. They concluded that many phishing emails use convincing company logos and do not contain spelling mistakes, such that those cues could not be relied on to identify a phishing attack.

Neither of these studies assessed whether these email cues are more or less likely to occur in genuine emails. Since users need to differentiate between phishing and genuine emails, it is vital to assess the cues of both, rather than focusing solely on phishing emails. These previous studies were also limited to the categorisation of email cues, and did not attempt to use this information to empirically determine which aspects of an email participants use to make legitimacy decisions. Our study attempts to address these issues.

## 2.2  Email cues identified by participants in previous studies

As highlighted in Section 1, a number of previous phishing studies have collected qualitative data on the email cues that participants used to make their legitimacy decisions.

For example, when Jakobsson (2007) asked participants to identify deceptive elements of phishing emails, the most important aspects they noted were spelling, grammar and design, followed by the source of the message, which included the link and reply-to address. Participants were more likely to trust emails with copyright information or legal disclaimers, but too much emphasis on security tended to decrease a participant's trust in an email (Jakobsson et al. 2007). In Furnell's (2007) study, participants were most influenced by visual factors (such as logos, banners, and copyright information), technical cues (such as the URL and use of 'https'), and language and content aspects (such as spelling mistakes, personalisation, and whether the message was overly urgent or forceful). Parsons et al. (2013) found that many participants were influenced by an inherent trust in the company from which the email appeared to originate. Other cues included visual presentation, spelling and grammatical errors, personalisation, and potential incentives within the email (Parsons et al., 2013). Similarly, Egelman et al. (2008) also found that participants were influenced by an inherent trust in the purported company and by personalisation. It was also reported that when participants looked at the URL, they did not always make an accurate decision regarding its authenticity. Many participants trusted phishing emails enough to visit the URLs and, in fact, many then failed to heed the active security warnings in the subsequent website (Egelman et al. 2008).

Complementary research (Fogg et al, 2003) has examined the cues of websites and asked participants to make comments about the credibility of each site. In this study, 46% of comments focused on the design and look of the website. Other comments included the structure and focus of information and the underlying motive of the website. Dhamija et al. (2006) found participants failed to utilise cues such as the address bar, status bar and security indicators, and were instead frequently influenced by visual presentation.

However, these findings are all based on participants' self-report of, rather than an empirical evaluation of, the cues that participants use to differentiate between phishing and genuine emails. In order to measure these factors empirically, it is first necessary to have a comprehensive understanding of the cues that can exist in phishing and real emails.

## 3  Stage One: Identifying the Cues that Differentiate Between Phishing and Genuine Emails

### 3.1  Stage One: Methodology

We used the findings reported in Section 2.2 to develop a list of cues that were commonly used to differentiate between phishing and genuine emails. The aim was to develop a list of cues that could be used to judge both types of emails. Any cue that was judged as important in any of the reviewed publications was added to the list. Content analysis was then used to group cues that fit in the same category. This resulted in 13 cues, which are listed in Table 1, together with the associated publications and the authors' description of the cues.

| Cue | Description used by authors | Source publication |
| --- | --- | --- |
| Consistency | Structure and focus of information | (Fogg et al., 2003) |
| Links | Source of the message: link | (Jakobsson, 2007) |
|  | Technical cues: URL | (Furnell, 2007) |
|  | Technical cues: https | (Furnell, 2007) |
|  | URL | (Egelman et al., 2008) |
|  | Address bar | (Dhamija et al., 2006) |
| Visual presentation | Logo | (Kim & Kim, 2013) |
|  | Logo | (Blythe et al., 2011) |
|  | Design | (Jakobsson, 2007) |
|  | Visual factors: logos | (Furnell, 2007) |
|  | Visual factors: banners | (Furnell, 2007) |
|  | Visual factors: copyright information | (Furnell, 2007) |

|  | Visual presentation | (Parsons et al., 2013) |
|---|---|---|
|  | Design and look | (Fogg et al., 2003) |
|  | Visual presentation | (Dhamija et al., 2006) |
| Personalisation | Language and content aspects: personalisation | (Furnell, 2007) |
|  | Personalisation | (Parsons et al., 2013) |
|  | Personalisation | (Egelman et al., 2008) |
| Spelling and grammatical errors | Spelling and grammar | (Blythe et al., 2011) |
|  | Spelling and grammar | (Jakobsson, 2007) |
|  | Language and content aspects: spelling mistakes | (Furnell, 2007) |
|  | Spelling and grammatical errors | (Parsons et al., 2013) |
| Security | Status bar | (Dhamija et al., 2006) |
|  | Security indicators | (Dhamija et al., 2006) |
|  | Security | (Jakobsson et al., 2007) |
| Legal | Copyright information or legal disclaimers | (Jakobsson et al., 2007) |
| Sender | Sender's address | (Kim & Kim, 2013) |
|  | Contact methods | (Kim & Kim, 2013) |
|  | Purported sender | (Blythe et al., 2011) |
|  | Source of the message: reply-to address | (Jakobsson, 2007) |
| Familiarity | Source credibility | (Kim & Kim, 2013) |
|  | Trust in purported company | (Parsons et al., 2013) |
|  | Inherent trust in company | (Egelman et al., 2008) |
| Importance | Rational appeals | (Kim & Kim, 2013) |
| Urgency | Time pressure | (Kim & Kim, 2013) |
|  | Language & content aspects: overly urgent or forceful | (Furnell, 2007) |
| Positive consequences | Potential incentives | (Parsons et al., 2013) |
|  | Emotional appeals | (Kim & Kim, 2013) |
|  | Motivational appeals | (Kim & Kim, 2013) |
|  | Premise of the email | (Blythe et al., 2011) |
|  | Underlying motive of the website | (Fogg et al., 2003) |
| Negative consequences | Emotional appeals | (Kim & Kim, 2013) |
|  | Motivational appeals | (Kim & Kim, 2013) |
|  | Premise of the email | (Blythe et al., 2011) |
|  | Underlying motive of the website | (Fogg et al., 2003) |

*Table 1: Email cues identified in previous research*

Using the 50 emails presented in Parsons et al. (2013), five participants rated each email on the presence or absence of the 13 cues listed in Table 1. These emails comprised 25 phishing and 25 genuine emails, and included emails from a range of topics, including banking, shopping and social networking emails. All participants had post-graduate qualifications; four were registered psychologists and one had a PhD in information security. They all have knowledge and skill in the topic and can therefore be considered 'Experts'. For each cue, an explanatory statement was developed. For the first 11 of the cues, experts were asked to respond to the following statements on a five-point scale from 'Strongly Disagree' to 'Strongly Agree':

    Consistency:         "The message within this email is consistent"
    Links:                   "The links within this email are legitimate"
    Visual presentation:  "The visual presentation of this email makes it appear legitimate"
    Personalisation:     "This email is personalised to the recipient"
    Spelling:              "This email has spelling and / or grammatical irregularities"
    Security:             "This email contains security advice"
    Legal:                 "The email contains copyright information and legal disclaimers"
    Sender:              "The email appears to be from the claimed sender"

    Familiarity:          "I am familiar with the named organisation or company"
    Important:          "This email is important"
    Urgency:           "This email is urgent"

Experts were also asked to rate the final two cues in regards to the possible positive consequences (e.g., gain) or negative consequences (e.g., threat) specified within each email. They could respond with 'N/A', 'Unclear / Not Sure', 'Low' or 'High'. For each cue, experts were provided with a definition to ensure they had a consistent understanding of the cue. For example, the following definitions were provided for consistency, visual presentation and email importance, respectively:

- *"This refers to whether the email has a conflicting or stable message throughout"*

- *"This refers to whether the email has a professional design, and should take into account whether the logos, colour schemes and other imagery seems consistent with the claimed sender"*

- *"This refers to the importance of the email, and should take into account whether the email could be of significance or value to the recipient"*

### 3.2  Stage One: Results

To determine the agreement between the ratings, participants' responses were assigned a numerical value (e.g., Strongly disagree = 1, Strongly agree = 5) and the inter-rater reliability was analysed using a two-way mixed, consistency, average-measures intra-class correlation coefficient (ICC) (McGraw et al. 1996; Shrout et al. 1979). Based on Cicchetti's (1994) guidelines, the resulting ICC was in the excellent range (*ICC* = .918, *CI* 95% = [.908, .927]), indicating a high degree of reliability between experts. Given this high degree of reliability, an average score was deemed to be suitable for use in further analysis.

To determine which of the cues rated by experts could be used to distinguish between phishing and genuine emails, a series of independent samples t-tests were conducted. This analysis determines whether any of the cues were significantly more likely to occur in either phishing or genuine emails. This analysis requires non-orthogonal planned comparisons, which means the comparisons of interest are not independent. To reduce the probability of Type 1 error caused by these 13 comparisons, the significance level (α) was adjusted to .004 using the Bonferroni technique (Dunn 1961). This approach ensures the family-wise alpha level remains at .05, and therefore also minimises the risk of Type 2 error. The results for phishing and genuine emails for all rated cues are displayed in Table 2.

|  | Type of email | | t-test | | |
|---|---|---|---|---|---|
|  | **Phishing** | **Genuine** | | | |
| **Cues** | M (SD) | M (SD) | t | df | *p* value |
| Consistency | 4.08 (0.56) | ***4.54 (0.39)*** | -3.427 | 48 | ***.001**** |
| Links | 2.01 (0.64) | ***4.16 (0.46)*** | -13.311 | 41.85 | ***.000**** |
| Visual presentation | 3.49 (1.04) | 3.90 (0.91) | -1.476 | 48 | .147 |
| Personalisation | 1.47 (0.79) | ***3.18 (1.30)*** | -5.629 | 39.67 | ***.000**** |
| Spelling | ***2.35 (1.03)*** | 1.62 (0.60) | 3.042 | 38.67 | ***.004**** |
| Security | 1.87 (0.71) | 2.14 (1.05) | -1.042 | 48 | .303 |
| Legal | 2.50 (1.31) | 2.50 (1.31) | -0.022 | 48 | .983 |
| Sender | 3.07 (0.98) | ***4.40 (0.50)*** | -6.037 | 35.88 | ***.000**** |
| Familiarity | 3.92 (0.98) | 3.93 (1.06) | -.028 | 48 | .978 |
| Importance | 2.38 (0.41) | 2.82 (0.77) | -2.522 | 36.84 | .016 |
| Urgency | 2.98 (0.76) | 2.44 (0.71) | 2.614 | 48 | .012 |
| Positive consequences | 1.49 (0.55) | 1.18 (0.58) | 1.943 | 48 | .058 |
| Negative consequences | 1.24 (0.83) | 0.70 (0.57) | 2.702 | 42.28 | .010 |

\* $p < .004$

*Table 2: Independent samples t-test for expert ratings of emails*

Results in Table 2 reveal that genuine emails were significantly more likely to be rated as consistent and personalised. Genuine emails were also significantly more likely than the phishing emails to have links and a purported sender that were rated to be legitimate. Phishing emails, on the other hand, were significantly more likely to contain spelling and grammatical irregularities.

The comparisons associated with positive and negative consequences, importance and urgency approached significance, but did not exceed the alpha level of interest ($\alpha = .004$). There were no significant differences between genuine and phishing emails based on visual presentation, the familiarity of the purported organisation, or the security advice, copyright information or legal disclaimers within the email. Hence, cues such as the visual presentation of the email, familiarity of the named organisation, and the security advice, copyright information or legal disclaimers within the email could not be used to differentiate between phishing and genuine emails.

# 4 Stage Two: Identifying the Cues that Participants Use

## 4.1 Stage Two: Methodology

To determine which of the email cues participants used to differentiate between phishing and genuine emails, the original data in Parsons et al.'s (2013) study was re-examined using the ratings reported in Section 3.2. In Parsons et al. (2013), 59 university students participated in a lab-based, role-play experiment and were asked to manage 50 emails[1], with all participants being asked to respond to the question, *"How would you manage this email?"* with one of the four replies:
   a) leave the email in the inbox and flag for follow up;
   b) leave the email in the inbox;
   c) delete the email;
   d) delete the email and block the sender.

For the purposes of the following analysis, a *genuine* email was deemed to be correctly managed if participants responded with options (a) or (b), and a *phishing* email was deemed to be correctly managed if participants responded with options (c) or (d). Although this scoring system may not always reflect participants' decisions on the legitimacy of an email, the validity of this approach for representing participants' intentions is demonstrated in Parsons et al. (2014).

The expert ratings were also converted to a binary score, so each cue was either 'present' or 'absent' in each email. For the cues assessed on a scale from 'Strongly Disagree' to 'Strongly Agree', the cue was deemed to be 'present' if the experts' consensus score was either 'Agree' or 'Strongly Agree', and 'absent' for all other response options (i.e., 'Strongly Disagree', 'Disagree' and 'Neither Agree Not Disagree'). For the other scale types, a cue was deemed to be 'present' if the consensus score was 'Low' or 'High', and 'absent' for all other response options (i.e., 'N/A' and 'Unclear / Not Sure').

After rescoring, 2 of the 13 cues were excluded from further analysis. Consistency was removed as only 2 of the 50 emails were judged to be inconsistent, and similarly, Security was removed as only 5 of the 50 emails were judged to contain security advice.

## 4.2 Stage Two: Results

For each participant, accuracy scores were calculated for both the presence and absence of each cue, where accuracy is the ability to discriminate between phishing and genuine emails. These scores determine whether participants' performance was influenced by any of the cues. A large difference in performance when a cue was present versus absent indicates that the cue may have been a determinant in participants' decision making. For example, one participant had an accuracy of only 17% for the emails where visual presentation was judged to be absent, and an accuracy of 61% for the emails where visual presentation was judged to be present. This suggests that this participant's performance was influenced by the visual presentation of an email; they tended to make good decisions for emails that had high visual appeal, and poor decisions for the emails without a professional design or colours and logos.

---

[1] These participants were not specifically told that they were taking part in a phishing study. A further 58 participants completed the same experiment and were primed to the phishing aspect of the study. Since the non-primed participants better represent the real world (where people are infrequently reminded of the risks of phishing) the data from these non-primed participants are re-examined in this study.

To evaluate the impact of the presence or absence of each cue on participants' performance, a series of paired-sample t-tests were conducted. Since this analysis requires non-orthogonal comparisons, the alpha level was adjusted to .005 using the Bonferroni technique to account for 11 non-orthogonal tests.

As shown in Table 3, these results indicate that several cues significantly influenced performance. Participants were significantly more likely to correctly manage an email when visual presentation was professional and when the email was personalised and important. They were also more accurate at judging emails that did not have spelling and grammatical errors or legal disclaimers, and discrimination performance was worse when emails had urgency or positive consequences. Participants were least likely to correctly manage emails that were urgent, with overall accuracy of only 42% for these emails. Since the results in Section 4.2 revealed that visual presentation, legal disclaimers, importance and urgency could not differentiate between phishing and genuine emails, this may mean that participants used poor cues to make legitimacy decisions.

|  | Occurrence of cue | | | | |
|---|---|---|---|---|---|
|  | **Absent** | **Present** | t-test | | |
| **Cues** | M (SD) | M (SD) | t | df | *p* value |
| Links | 0.46 (0.21) | 0.58 (0.20) | -2.50 | 58 | .015 |
| Visual presentation | 0.44 (0.17) | ***0.54 (0.09)*** | -4.83 | 58 | ***.000*** |
| Personalisation | 0.44 (0.14) | ***0.67 (0.17)*** | -6.93 | 58 | ***.000*** |
| Spelling | ***0.55 (0.10)*** | 0.43 (0.17) | 4.99 | 58 | ***.000*** |
| Legal | ***0.55 (0.10)*** | 0.45 (0.12) | 6.67 | 58 | ***.000*** |
| Sender | 0.50 (0.22) | 0.52 (0.22) | -.49 | 58 | .623 |
| Familiarity | 0.51 (0.14) | 0.52 (0.10) | -.07 | 58 | .941 |
| Importance | 0.46 (0.11) | ***0.75 (0.16)*** | -11.37 | 58 | ***.000*** |
| Urgency | ***0.55 (0.09)*** | 0.42 (0.16) | 6.74 | 58 | ***.000*** |
| Positive consequences | ***0.55 (0.13)*** | 0.50 (0.09) | 3.03 | 58 | ***.004*** |
| Negative consequences | 0.53 (0.10) | 0.49 (0.16) | 1.65 | 58 | .104 |

*Table 3: Paired samples t-test for proportion correct based on email cues*

## 5   Discussion

This study presented a detailed categorisation of the cues of 25 phishing and 25 genuine emails. Based on the judgements of five experts, findings suggest that consistency, links, personalisation, spelling and the sender are the best indicators of the genuineness of an email. Generally speaking, genuine emails are more likely to have a stable message throughout, more likely to have links that appeared legitimate, more likely to be addressed to the recipient, are less likely to have spelling or grammatical errors, and are more likely to be from a sender who appeared legitimate.

Our results demonstrated that participants often make decisions based on poor indicators of the genuineness of an email. For example, participants were influenced by visual presentation; they tended to make more accurate decisions when faced with emails with a professional looking design or logo, but poor decisions for emails with poor visual presentation.

Results also indicated that participants were influenced by the urgency of an email. The lowest performance for any category occurred when participants judged emails that were rated as urgent by the experts. This is particularly concerning from a practical point of view. Essentially, evidence suggests that the average phishing site is taken down after approximately 62 hours (Hong 2012), which means phishing emails require victims to make a quick decision. From a practical point of view, users who are influenced by urgency are more likely to be phished. Hence, this study highlights a need to educate participants on the dangers associated with a quick or urgent response to phishing emails.

This study has implications for education and training and provides a basis for the design and development of targeted and more relevant training and risk communication strategies. It indicates which cues were the best indicators of a genuine email, and also which cues influenced participants. This highlights the areas where education should be focused to change behaviour. For example, participants were significantly more accurate when viewing emails without legal disclaimers. This means the presence of copyright information or legal disclaimers degraded a participant's ability to accurately assess an email. Training and education should emphasise that certain cues, such as a legal disclaimer, can be copied by people with malicious intentions, and are therefore poor indicators of an email's legitimacy. Similarly, training and education could highlight the ease with which an email can be created to have a high visual representation due to the ready access of corporate logos etc. via electronic means. Furthermore, although certain cues are good indicators of an email's legitimacy, education should emphasise that these cues are indicative and not absolute or infallible. For example, it is common to find poor spelling or grammar in genuine emails, and not all phishing emails contain mistakes. In addition, since scammers continually improve the sophistication of phishing emails (Arachchilage et al. 2013; Arachchilage et al. 2014; Purkait 2012), it is important that any training is regularly updated to reflect current trends.

These findings also have important implications for marketing and business. Companies who want to ensure their clients or customers will trust their emails should use quality visual presentation and should not emphasise the urgency or positive consequences associated with their email. Instead, they should use personalisation and should emphasise the importance of the email.

### 5.1 Limitations and Future Directions

This study used a selection of actual phishing and genuine emails such that the effect of the cues in emails was not controlled. Hence, in each email, a number of different cues occurred simultaneously, and some cues did not occur with equal frequency for phishing and genuine emails. For example, of the 17 emails that were judged to be 'personalised', only 2 were phishing emails. In contrast, 23 phishing emails and 10 genuine emails were not personalised. Hence, correctly managing emails without personalisation was a more difficult task than managing personalised emails. Consequently, the results of this study do not reveal the influence of a single cue on participants' performance. In a future study, the presence or absence of each cue could be more strictly controlled.

It is also important to note that participants were not required to act; the options only required them to identify how they *would* act. Although previous research has proven that the categories utilised provide a reasonable representation of the intentions of participants (Parsons et al., 2014), results are solely based on a lab-based, role-play experiment. It is also important to note that the role-play experiment limits participants' ability to make context dependent decisions. For example, participants did not know whether the recipient subscribes to a particular bank or social networking site. Ideally future research could obtain permission and ethics approval to monitor how individuals make decisions for emails received in their own inboxes to better understand which aspects of an email are most influential.

In addition, future research could calculate the presence or absence of cues using non-binary criteria. This may reveal whether participants' performance varied based on the level at which each cue was present. Future research could also replicate this study using a larger number of experts. This could verify whether the cues identified by the five experts used in this study reflect those identified by a larger expert group. Finally, the cues judged by experts in this research were based on the cues identified in previous research, and few of these were based on the social engineering techniques that may compel an individual to respond. Future research could assess the influence of social engineering techniques like authority or social influence to determine which of these most influence different types of individuals.

## 6 Conclusion

In conclusion, the research of this paper expands on the largely speculative nature of research regarding the cues that participants use to differentiate between a phishing and genuine email. It was found that compared to genuine emails, phishing emails are more likely than genuine emails to have spelling or grammatical errors and are less likely than genuine emails to be personalised, consistent, and less likely to have links and a sender that appears legitimate. This information could form the basis of phishing training and education.

This paper also revealed that participants' legitimacy decisions are often influenced by incorrect cues. For example, participants were influenced by more extraneous aspects of an email, such as visual presentation and the positive consequences in the email. This suggests that people require further training and education on the aspects of an email that are the best indicators of the genuineness of an email. However, since phishing emails are becoming increasingly sophisticated, it is important that researchers and organisations do not become complacent, and instead, regularly review the cues of both phishing and genuine emails, and the cues that influence user performance, to assist users to make their decisions based on the most useful information.

# 7　References

Arachchilage, N. A. G., and Love, S. 2013. "A game design framework for avoiding phishing attacks," *Computers in Human Behavior* (29:3), pp 706-714.

Arachchilage, N. A. G., and Love, S. 2014. "Security awareness of computer users: A phishing threat avoidance perspective," *Computers in Human Behavior* (38), pp 304-312.

Blythe, M., Petrie, H., and Clark, J. A. Year. "F for fake: four studies on how we fall for phish," Proceedings of the SIGCHI Conference on Human Factors in Computing Systems ACM, Vancouver, Canada., 2011, pp. 3469-3478.

Cicchetti, D. V. 1994. "Guidelines, criteria, and rules of thumb for evaluating normed and standardized assessment instruments in psychology," *Psychological Assessment* (6:4), p 284.

Dhamija, R., Tygar, J. D., and Hearst, M. Year. "Why phishing works," Proceedings of the SIGCHI conference on Human Factors in computing systems, ACM2006, pp. 581-590.

Dunn, O. J. 1961. "Multiple comparisons among means," *Journal of the American Statistical Association* (56:293), pp 52-64.

Egelman, S., Cranor, L. F., and Hong, J. Year. "You've been warned: an empirical study of the effectiveness of web browser phishing warnings," Proceedings of the SIGCHI Conference on Human Factors in Computing Systems, ACM2008, pp. 1065-1074.

Ferguson, A. J. 2005. "Fostering e-mail security awareness: The West Point carronade," *EDUCASE Quarterly* (28:1), pp 54-57.

Furnell, S. 2007. "Phishing: can we spot the signs?," *Computer Fraud & Security* (2007:3), pp 10-15.

Furnell, S. 2013. "Still on the hook: the persistent problem of phishing," *Computer Fraud & Security* (2013:10), pp 7-12.

Hong, J. 2012. "The state of phishing attacks," *Communications of the ACM* (55:1), pp 74-81.

Jagatic, T. N., Johnson, N. A., Jakobsson, M., and Menczer, F. 2007. "Social phishing," *Communications of the ACM* (50:10), pp 94-100.

Jakobsson, M. 2007. "The human factor in phishing," *Privacy & Security of Consumer Information* (7), pp 1-19.

Jakobsson, M., Tsow, A., Shah, A., Blevis, E., and Lim, Y.-K. 2007. "What instills trust? a qualitative study of phishing," in *Financial Cryptography and Data Security*, Springer, pp. 356-361.

Kim, D., and Kim, J. H. 2013. "Understanding persuasive elements in phishing emails: a categorical content and semantic network analysis," *Online Information Review* (37:6), pp 2-2.

McGraw, K. O., and Wong, S. 1996. "Forming inferences about some intraclass correlation coefficients," *Psychological Methods* (1:1), p 30.

Olivo, C. K., Santin, A. O., and Oliveira, L. S. 2013. "Obtaining the threat model for e-mail phishing," *Applied Soft Computing* (13:12), pp 4841-4848.

Osterman Research Inc. 2015. "Best Practices for Dealing with Phishing and Next-Generation Malware: An Osterman Research White Paper," Black Diamond, Washington.

Parsons, K., McCormac, A., Pattinson, M., Butavicius, M., and Jerram, C. Year. "Phishing for the truth: A scenario-based experiment of users' behavioural response to emails," Security and Privacy Protection in Information Processing Systems - IFIP Advances in Information and Communication Technology, Springer2013, pp. 366-378.

Parsons, K., McCormac, A., Pattinson, M., Butavicius, M., and Jerram, C. Year. "Using Actions and Intentions to Evaluate Categorical Responses to Phishing and Genuine Emails," Proceedings of the Eighth International Symposium on Human Aspects of Information Security and Assurance (HAISA 2014), University of Plymouth, Plymouth, UK, 2014.

Pattinson, M., Jerram, C., Parsons, K., McCormac, A., and Butavicius, M. 2012. "Why do some people manage phishing emails better than others?," *Information Management & Computer Security* (20:1), pp 18-28.

Purkait, S. 2012. "Phishing counter measures and their effectiveness–literature review," *Information Management & Computer Security* (20:5), pp 382-420.

Shrout, P. E., and Fleiss, J. L. 1979. "Intraclass correlations: uses in assessing rater reliability," *Psychological Bulletin* (86:2), p 420.

## Acknowledgements

This project was supported by a Premier's Research and Industry Fund grant provided by the South Australian Government Department of State Development.